\documentclass[conference]{IEEEtran}
\usepackage{color}
\usepackage{multicol}
\usepackage{graphicx}
\usepackage{multirow}
\usepackage{booktabs}
\usepackage[table]{xcolor}

\ifCLASSINFOpdf

\else
\fi

\begin{document}
%
\title{Adaptive-Reliable Medium Access Control\\ Protocol for Wireless Body Area Networks}

\author{\IEEEauthorblockN{A. Rahim, N. Javaid, M. Aslam, U. Qasim$^{\ddag}$, Z. A. Khan$^{\S}$\\}

        $^{\ddag}$University of Alberta, Alberta, Canada\\
        Department of Electrical Engineering, COMSATS\\ Institute of
        Information Technology, Islamabad, Pakistan. \\
        $^{\S}$Faculty of Engineering, Dalhousie University, Halifax, Canada.
        }

\maketitle

\section{Motivation}

 Extensive energy is consumed by Transceiver communication operation [1].  Existing research on MAC layer focuses to maximize battery-powered sensor node's life. Bottleneck of MAC layer protocol design for WBAN is to achieve high reliability and energy minimization. Majority of MAC protocols designed for WBANs are based upon TDMA approach. However, a new protocol needs to be defined to achieve high energy efficiency, fairness and avoid extra energy consumption due to synchronization.

\section{Protocol design}
Proposed MAC protocol, AR-MAC is based upon TDMA approach to minimize energy consumption. AR-MAC assigns Guaranteed Times Slot (GTS) to each sensor node for communication based upon the requirements of sensor node . To reduce overhearing and idle listening, proposed system uses periodic sleep and wakeup according to node requirements. We assume a star topology; a Central Node (CN) collects data from sensor nodes and communicates with a Monitoring Station (MS), direct or through an Access Point (AP).

%


CN is usually equipped with larger batteries and higher computational power. One or two transceivers may be used within a single CN. In case of two transceivers total time frame $T^{Frame}$ is allocated for communication with sensor nodes.  We assume CN with single transceiver where $T^{Frame}$  is divided into three parts: Contention Free Period (CFP) for communication with sensors, Contention Access Period (CAP) to accommodate emergency or on-demand traffic and time $T^{MS}$ for communicating sensor nodes' data to MS.

\subsection{Channel Selection}

Initially, CN starts scanning for available free Radio Frequency (RF) channels. If the current RF Channel is busy, CN switches to another RF Channel. CN selects a free RF channel for communication. After successful selection of RF Channel, CN broadcasts the Channel Packet with address and channel information to sensor nodes. On the other side end nodes scan RF channels for Channel Packet from CN. Sensor node scans the RF channel if it is free it switches to another RF channel. If the channel is busy it waits for time $T^{CP}$ to listen Channel Packet. If sensor node does not receive the Channel Packet, it switches again to next channel. After successful reception of Channel Packet, node starts transmission and sends an acknowledgment (ACK) packet to CN.

\subsection{Time Slot Assignment}
Once sensor node selects a proper RF Channel after receiving Channel Packet from CN, sensor node sends out a Time Slot Request (TSR) packet to CN. TSR packet includes sensor node's data rate and required time slot information. Authors in [6], propose time slots of fixed length with fixed guard band time, $T^{GB}$. In-body and on-body biomedical sensors have different data rates and sampling intervals with different clock drifts. Assigning time slots of equal length for unequal requirements is wastage of resources. AR-MAC uses an adaptive scheme for Time Slot (TS) and GBT time. Based on traffic pattern of nodes, CN assigns time slot and sends Time Slot Request Reply (TSRR). These time slots are of variable length depend upon the requirements of sensor nodes. Assigned time slot can easily accommodate the transmission of data packet, reception of ACK packet and some acceptable delay, based on the communication model. $T^{GB}$ is inserted between the two successive time slots to avoid the interference due to clock drift of node and CN as shown in Fig.3.
\begin{figure}[ht]
\begin{center}
\includegraphics[scale=0.35]{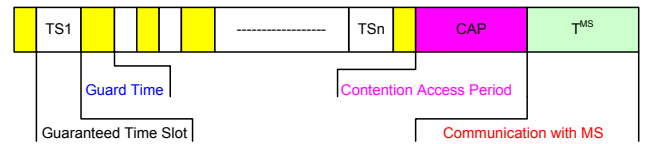}
\caption{Time Slots Assignment with Guard-band Time}
\end{center}
\end{figure}

Value of $T^{GB}$  depends upon length of the successive time slots. Adaptive GB avoids possibilities of collision and interference due to clock drift. We calculate $T^{GB}$ as follows:

\begin{eqnarray}
T^{GB}_{n,n+1} = \frac{F}{100} \times \frac{1}{2}[ TS_n + TS_{n+1}]
\end{eqnarray}

\begin{eqnarray}
T^{GB}_1 = \frac{F \times TS_1}{100}
   \end{eqnarray}

\begin{eqnarray}
       T^{GB}_n = \frac{F \times TS_n}{100}
    \end{eqnarray}
Where $F$ is guard band factor, depends upon the average drift value.  However guard band time $T^{GB}_1$ is inserted before first time slot and similarly guard band time $T^{GB}_n$ is placed after $N$ time slot. After successful time slot assignment, sensor nodes enter into sleep mode and wakeup only to send data to CN in allocated time slots. Periodic sleep reduces energy consumption due to idle listening and Overhearing. Allocated time slots in CFP are completely collision free thus reduce the energy consumption and make the communication reliable.

\subsection{Synchronization}
TDMA schemes require extra energy cost for periodic synchronization [8]. Synchronization of nodes after N number of cycles is energy consuming process. AR-MAC uses a novel synchronization mechanism to avoid collision and energy consumption. After successful assignment of time slots to nodes, CN listens to data packet within expected time slot. Upon arrival of data packet, CN compares current arrival time of the packet and expected arrival time with acceptable delay $(D)$. Based on the difference of current arrival time and expected arrival time a Drift Value $(DV)$ is calculated. This $DV$ is  transmitted to node within ACKnowledgment (ACK) packet to adjust time slot for future communication. However, this value depends upon acceptable delay and $F$. If the difference between the expected arrival time and current arrival is greater than $D$, CN sends $DV$ with in SYNChronization ACKnowledgment (SYNC-ACK) packet for future synchronization to sensor nodes otherwise, CN sends simple ACK packet for received data packet. For communication of data in future, sensor node adjusts its wakeup time schedule according to $DV$. Using this scheme of synchronization a node can go into sleep mode without loosing synchronization for N number of cycles. Acceptable delay $D$ is linked with guard band factor $F$ as under:

\begin{eqnarray}
D = Min(TS_1 ...... TS_n) \times \frac {F}{100}
\end{eqnarray}

 For future synchronization, decision of sending $DV$ to end nodes is based upon the difference of current arrival time and expected arrival time of data packet.  $\bigtriangleup T$ represents this difference.

\begin{eqnarray}
    \bigtriangleup T = Expected Arrival Time - Current Arrival Time
\end{eqnarray}

\begin{eqnarray}
  DV = \left\{
  \begin{array}{l l}
    0 & \quad \textrm{if $\mid\bigtriangleup T\mid < D $ }\\
    \bigtriangleup T & \quad \textrm{if $\mid\bigtriangleup T\mid > D $}\\
  \end{array} \right.
\end{eqnarray}

\subsection{Frame Formate}
Proposed AR-MAC uses two types of packets: Data Packets and Control Packets. In data packet sensor node sends its periodic data in allocated time slot. For emergency data, node uses CAP. Control Packets are:
\begin{enumerate}

\item Channel Packet: \emph{After channel selection central node advertises channel information and its unique address in Channel Packet.}
\item Time Slot Request (RSR) Packet: \emph{Sensor node sends information to Central Node for Guaranteed Time Slot Assignment in Time Slot Request packet.}
\item Time Slot Request Reply (TSRR) Packet: \emph{Central node sends Guaranteed Time Slot information with CAP information to node in Time Slot Request Reply packet.}
\item Synchronization-Acknowledgment (SYNC-ACK) Packet: \emph{For synchronization, Central node sends the required Drift Value to end node with ACK of previously received data packet in Synchronization Packet.}
\item Data Request (DR) Packet: \emph{For on demand traffic/information, Central Node sends Data Request Packet to end  node.}
\item Acknowledgment (ACK) Packet \emph{Each data packet is acknowledged using Acknowledgment Packet }
\end{enumerate}

\begin{figure}[ht]
\begin{center}
\includegraphics[scale=0.6]{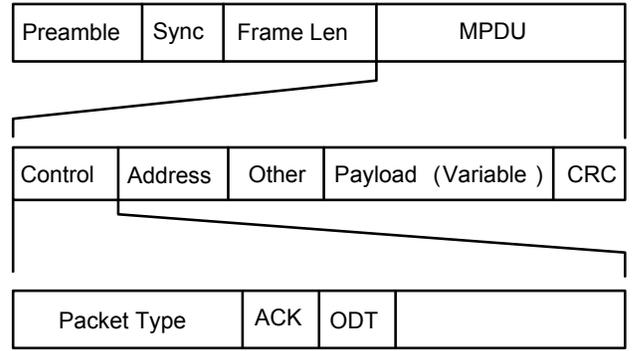}
\caption{MAC Layer Frame Formate}
\end{center}
\end{figure}

The MAC Protocol Data Unit (MPDU) stars with 3 octets of overhead. First octet carries information about packet type. Preceding two octets carry address information. In case of emergency traffic sensor node waits for CAP. Upon successful Clear Channel Assessment (CCA), sensor node starts communication with CN. However, in case of on-demand traffic CN sets On-Demand Traffic (ODT) field in ACK or SYNC-ACK packet to 1. After receiving this packet sensor node wakes up in CAP to listen data request from CN. After receiving data request from CN, sensor node sends requested data packets and waits for ACK packet. Sensor node enters into sleep mode after successful reception of ACK.

\section{Energy Consumption Analysis}
In order to model energy consumption, we consider the energy consumption related to transceiver. In this study, we assume energy consumption of sensing and processing units to be constant. We assume periodic traffic pattern, i.e., sensor nodes send periodic data to CN in assigned time slots. Most of the time sensor nodes remain in sleep mode. During allocated time slots, they wakeup to send data. We use the following equation to measure the energy consumption for $N$ number of cycles.
\begin{eqnarray}
E_{Total} =\sum_{k=1}^N E_{{Sleep}_k} + \sum_{k=1}^N E_{{Active}_k}
\end{eqnarray}

Energy consumption is a function of time and current drawing from voltage source for a specific task. When nodes enter into sleep mode they still consume energy. The sleep mode duration can be calculated from total time frame length and time for which the node is in active mode.

\begin{eqnarray}
T_{Sleep}=T_{Frame}-T_{Active}
\end{eqnarray}
\begin{eqnarray}
E_{Sleep}=T_{Sleep}\times I_{Sleep} \times V
\end{eqnarray}

$I_{Sleep}$ is the current drawing from voltage source $V$ during sleep mode. In $T_{Active}$ the nodes receive, transmit and wait for Acknowledgment. Energy is also consumed in switching, from sleep to active and active to sleep mode. The energy consumed for all these tasks will be considered as energy in $T_{Active}$.

\begin{eqnarray}
E_{Active}= 2 \times E_{Sw}+E_{Trans}+ E_{Rec}+E_{TOut}
\end{eqnarray}

Where $E_{Sw}$ is Switching energy, $E_{Trans}$ Transmission energy, $E_{Rec}$ is Receiving energy and $E_{TOut}$ is Time-Out energy. We describe these terms in details in the following subsections.

\subsection{Switching Energy}
Most of the time, sensor nodes remain in sleep mode. Sensor nodes turn on its transceiver in wakeup mode for communication. Switching energy is the consumed energy for switching transceiver between states; sleep mode and wakeup mode. Frequently switching of transceiver between states leads to high energy consumption. Energy consumed for switching the transceiver is determined by the following equation.

\begin{eqnarray}
E_{Sw} = T_{Switch} \times I_{Switch} \times V
\end{eqnarray}

Where $T_{Switch}$ is the required for the transceiver to switch between sleep and wakup mode and $I_{Switch}$ is the required current.

\subsection{Transmission Energy}

Transmission energy is the energy consumed for transmission of Data or Control packet of length $P$. Following equation links the transmission energy with length of packet $P$, time required for transmission of single byte $T_{Byte}$, current draw during transmission $I_{Trans}$ and a voltage source V.

\begin{eqnarray}
E_{Trans} = P \times T_{Byte} \times I_{Trans} \times V
\end{eqnarray}

\subsection{Receiving  Energy}
Receiving Energy is the consumed energy while receiving packets and their associated overhead. Receiving energy is expressed as:

\begin{eqnarray}
E_{Rec} = P \times T_{Byte} \times I_{Rec} \times V
\end{eqnarray}

Where $I_{Rec}$ is the Current during reception, $T_{Byte}$ is the time for single byte, $P$ is the length of packet and $V$ is the voltage source.

\subsection{Time-Out Energy}
The energy consumed after transmission and before reception of an ACK packet is termed as Time-Out energy. For time $T_{TOut}$, current $I_{TOut}$ and voltage source $V$, we used the equation given below to calculate the energy consumption during Time-Out.
\begin{eqnarray}
E_{T-Out} =  T_{TOut} \times I_{TOut} \times V
\end{eqnarray}

\section{Simulation Results}
We use MATLAB to measure and compare the  energy efficiency of the AR-MAC with that of IEEE 802.15.4. In energy consumption comparison, we consider the energy consumption of RF transceiver. We use the energy consumption model from  Crossbow MICAz data sheet as shown in Table II.   Packets are dropped randomly with average Packet Error Rate probability from 1\% to 20\%. Time frame size used in simulations is  $T^{Frame}$ =  1 Second. We used packets format as shown in Fig.4. Simulation has been carried out for 10 Sensor Nodes.
\vspace{-0.3cm}
\begin{table}[htbp]
  \centering
    \begin{tabular}{ll}
    \multicolumn{2}{c}{Table.1. Simulation Parameters Valu} \\
    \hline
    \toprule
    Parameter & Value  \\
    \midrule
    Time frame( $T^{Frame}$) & 1 Second  \\
    Voltage Source  & 3 volts \\
    Current Draw in Receive Mode  & 19.7 mA \\
    Current Draw in Transmit Mode & 17.4 mA   \\
    Current Draw in Idle Mode  & 20.0 mA   \\
    Current Draw in Sleep Mode  & 1 micro-A   \\
    Number of Sensor Nodes  & 10   \\
    Number of Cycles $N$    &   1000\\
    \bottomrule
    \end{tabular}%
  \label{tab:addlabel}%
\end{table}%
\vspace{-0.5cm}
 \begin{figure}[ht]
\begin{center}
\includegraphics[scale=0.45]{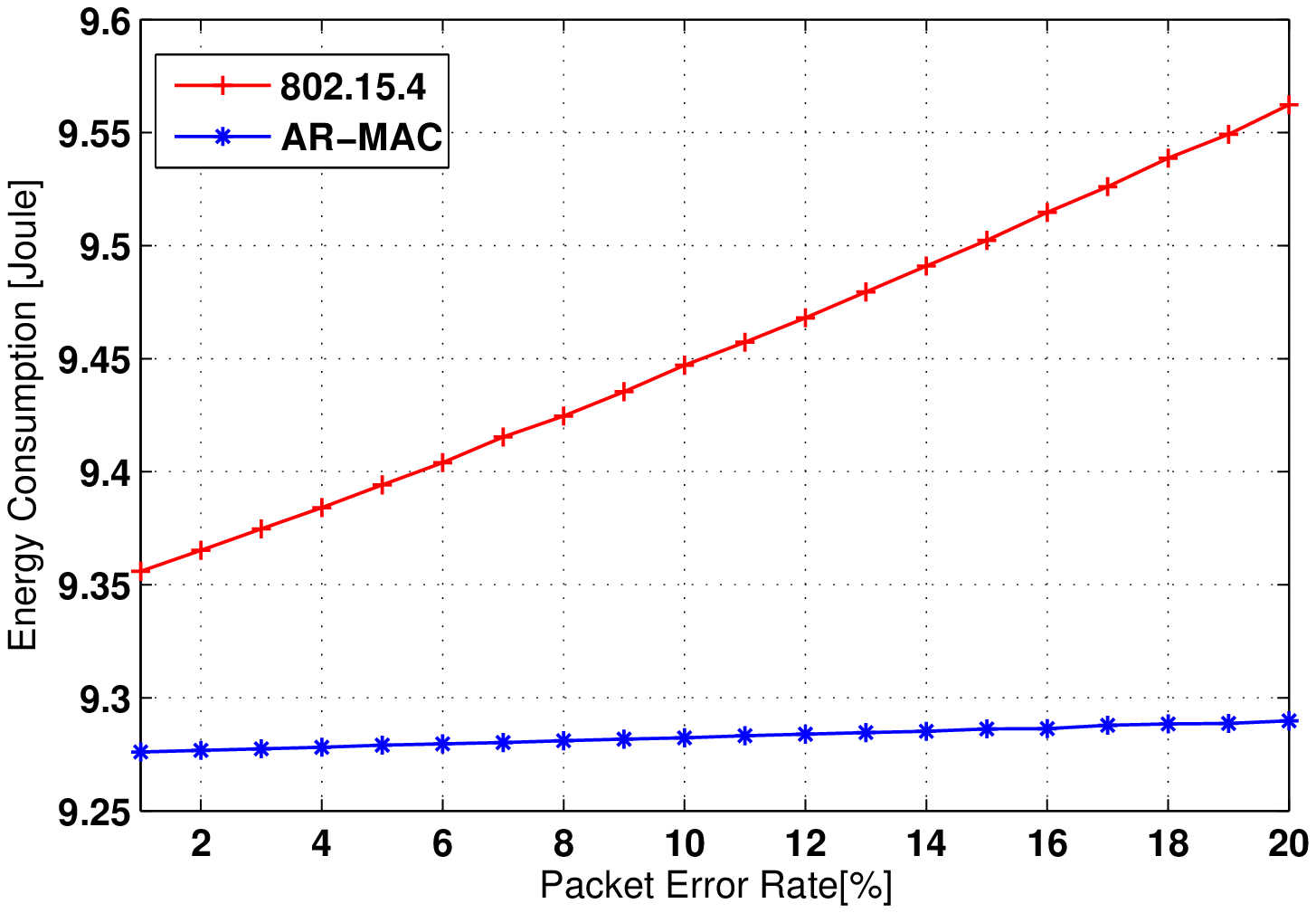}
\vspace{-0.5cm}
\caption{Energy consumption of AR-MAC and IEEE 802.15.4 for $N$ = 1000 }
\end{center}
\end{figure}
\vspace{-0.5cm}
We used Eq. 7 to calculate the energy consumption for $N$ = 1000. Figure 5 shows the energy comparison of the AR-MAC with IEEE 802.15.4.
The graph in Figure 5 shows that energy consumption of IEEE 802.15.4 increases with increase in probability of Packet Error Rate. This increase in energy consumption is due to extra energy requirement of CSMA/CA operation in IEEE 802.15.4. The energy consumption of AR-MAC increases with a minor variation due its adaptive time allocation and adaptive guard band mechanism. AR-MAC assignees guaranteed time slots to sensor nodes for communication, to overcome the packet collision and overhearing.


%


\end{document}